\documentclass[12pt]{article}
\usepackage{amsmath}
\usepackage{graphicx}
\usepackage{subfigure}
\usepackage{hyperref}

\begin{document}

\author{Ernst Trojan \and \textit{Moscow Institute of Physics and Technology} \and 
\textit{PO Box 3, Moscow, 125080, Russia}}
\title{Superluminal self-interacting neutrino}
\maketitle

\begin{abstract}
The effect of nonlinear self-interaction can be associated with superluminal
velocity of neutrino. The power energy spectrum $E=p+Cp^a$ is derived from
the nonlinear Dirac equation when interaction term $V=\lambda (\bar \psi
\gamma _\mu \psi \bar \psi \gamma ^\mu \psi )^a$ is added to the Lagrangian
of a free spin-1/2 particle. The superluminal velocity recorded by the OPERA
and MINOS collaborations is achieved when the coupling constants are taken
in the range $a=0.4\div 1.18$ and $\lambda =-\left( 0.5\div 1.6\right)
\times 10^{-4}$. The self-interaction Lagrangian $V=\lambda \bar \psi \gamma
_\mu \psi \bar \psi \gamma ^\mu \psi $ with the coupling constant $\lambda
=-\left( 0.7\div 0.9\right) \times 10^{-4}$ yields the same result. Scalar
interaction $V=\lambda (\bar \psi \psi )^b$ and scalar-vector interaction $%
\lambda \left( \psi ^{\dagger }\psi \right) ^{b+1}/\left( \bar \psi \psi
\right) ^b$ cannot be responsible for the observed superluminal neutrino.
\end{abstract}

\section{Introduction}

Neutrino was believed to be a massless spin-1/2 fermion with energy 
\begin{equation}
E=pc  \label{uu}
\end{equation}
and group velocity $v=c$ equal to the speed of light $c=1$ (in relativistic
units). The modern theory expects, however, that neutrino has finite mass 
\cite{M2011} 
\begin{equation}
m=m_\nu <0.28\,\mathrm{eV}  \label{mu}
\end{equation}
that implies deviation from the energy spectrum (\ref{uu}) and velocity 
\begin{equation}
v=\frac{dE}{dp}\neq 1  \label{v0}
\end{equation}
Recent experiments of the OPERA Collaboration \cite{OP} revealed
superluminal motion of neutrino with energy $E=17\,\mathrm{GeV}$ at the
average velocity 
\begin{equation}
v=1+\mathbf{2.37}\,\mathbf{\times }10^{-5}  \label{del}
\end{equation}
while the the MINOS Collaboration detected velocity 
\begin{equation}
v=1+\mathbf{5.1\times }10^{-5}  \label{del2}
\end{equation}
for the low energy neutrino with energy spectrum peaked at approximately $%
E=3\,\mathrm{GeV}$. Superluminal neutrino was also observed in supernova
explosion SN1987a \cite{SN87}.

This fact is a serious puzzle to the researchers. Superluminal velocity (\ref
{del}) cannot belong to a free massive particle with energy spectrum $E=%
\sqrt{p^2+m^2}$ whose velocity is always subluminal because 
\begin{equation}
v-1=-\frac 12\frac{m^2}{p^2}\simeq -\frac 12\frac{m^2}{E^2}<0  \label{vv0}
\end{equation}
The tachyonic energy spectrum $E=\sqrt{p^2-m^2}$ results in velocity above
the speed of light 
\begin{equation}
v-1=\frac 12\frac{m^2}{p^2}\simeq \frac 12\frac{m^2}{E^2}>0  \label{vv}
\end{equation}
that does not exceed $3\times 10^{-22}$ even at maximum possible neutrino
mass $m$ (\ref{mu}) and $E=17\,\mathrm{GeV}$. Nevertheless, in the frames of
the accuracy of measurements, the energy spectrum of superluminal neutrino
of OPERA \cite{OP} and MINOS \cite{MI} can be fitted to a power law \cite
{P1,P2,P3} 
\begin{equation}
E=p+Cp^a  \label{s}
\end{equation}
where the coefficients must be taken in the range \cite{T12} 
\begin{equation}
a=0.40\div 1.18\qquad C=4.15\times 10^{-4}\div 1.5\times 10^{-5}  \label{sa}
\end{equation}
and $C=\left( 2.2\,\div 3.03\right) {\times }10^{-5}$ if we choose $a\equiv 1
$.\textrm{\ }Indeed, neutrino is not a free particle, but there are several
interesting hypotheses to explain its superluminal motion \cite{SLN1}.

In the present paper we consider massive neutrino whose Lagrangian 
\begin{equation}
L=\bar \psi \left( i\gamma ^\mu \partial _\mu -m\right) \psi +V\left( \bar
\psi ,\psi \right)  \label{li}
\end{equation}
includes nonlinear self-interaction term $V\left( \bar \psi ,\psi \right) $.
There is no additional interaction with external fields and the medium,
while the superluminal velocity is hidden in the very nature of neutrino. We
need to find the energy spectrum of nonlinear Dirac equation 
\begin{equation}
\left( i\gamma ^\mu \partial _\mu -m\right) \psi +\frac{\partial V}{\partial
\bar \psi }=0  \label{d0}
\end{equation}
and check whether the Lagrangian (\ref{li}) is adjustable to reproduce the
superluminal velocity (\ref{del})-(\ref{del2}) detected by the OPERA \cite
{OP} and MINOS \cite{MI} collaborations.

\section{Effect of self-interaction}

Consider the Lagrangian (\ref{li}) with a simple self-interaction term 
\begin{equation}
V\left( \bar \psi ,\psi \right) =\frac \lambda {b+1}\left( \psi ^{\dagger
}\psi \right) ^{b+1}\qquad \psi ^{\dagger }=\bar \psi \gamma ^0  \label{u}
\end{equation}
The relevant Dirac equation (\ref{d0}) is written 
\begin{equation}
\left( i\gamma ^\mu \partial _\mu -m+F\right) \psi =0  \label{d}
\end{equation}
where 
\begin{equation}
F=\gamma ^0\omega   \label{fg}
\end{equation}
and 
\begin{equation}
\omega =\lambda \left( \psi ^{\dagger }\psi \right) ^b  \label{f}
\end{equation}
The Dirac equation (\ref{d}) at $\omega =0$ has well-known solution in the
form of plane wave 
\begin{equation}
\psi _0=\left( 
\begin{array}{c}
\phi _0 \\ 
\chi _0
\end{array}
\right) \exp \left( -iE_0t\right) \qquad \chi _0=\frac{\vec \sigma \cdot
\vec p}{E_0+m}\phi _0\qquad E_0=\sqrt{m^2+p^2}  \label{om0}
\end{equation}

Substituting stationary wave function 
\begin{equation}
\psi =\varphi \left( \vec r\right) \exp \left( -iEt\right)  \label{om}
\end{equation}
in (\ref{d}), we have equation 
\begin{equation}
\left( -i\vec \alpha \cdot \nabla +\beta m\right) \varphi =\left( E+\omega
\right) \varphi  \label{db}
\end{equation}
where 
\begin{equation}
\vec \alpha =\beta \vec \gamma \qquad \beta =\gamma ^0=-\gamma _0
\label{alf}
\end{equation}
Substituting a plane-wave bispinor 
\begin{equation}
\varphi =\left( 
\begin{array}{c}
\phi \\ 
\chi
\end{array}
\right) \exp \left( i\vec p\cdot \vec r\right)  \label{omb}
\end{equation}
in (\ref{db}), we obtain a linear system of equations 
\begin{equation}
\begin{array}{c}
\vec \sigma \cdot \vec p\chi =\left( E+\omega -m\right) \phi \\ 
\vec \sigma \cdot \vec p\phi =\left( E+\omega +m\right) \chi
\end{array}
\label{dc}
\end{equation}
that has solution if and only if 
\begin{equation}
E=\sqrt{m^2+p^2}-\omega  \label{e0}
\end{equation}
where $p=\left| \vec p\right| $.

We can estimate the energy spectrum (\ref{e0}) in the frames of mean-field
approximation if we neglect correlations between the field operators in (\ref
{f}) and apply effective interaction 
\begin{equation}
\omega \simeq \omega _{*}=\lambda n^b  \label{fb}
\end{equation}
where 
\begin{equation}
n=\left\langle \psi ^{\dagger }\psi \right\rangle  \label{n}
\end{equation}
is the particle number density. The latter can be adjusted as $n=1/V$ for 1
particle in volume $V$. For a many-particle system the quantity (\ref{n}) is
determined according to formula

\begin{equation}
n=\frac 2{\left( 2\pi \right) ^q}\int\limits_0^\infty f_k\,d^qk  \label{n1}
\end{equation}
where $f_p$ is the distribution function in $q$-dimensional momentum space.
The latter is the Fermi-Dirac distribution function if it is an ideal gas in
equilibrium, however, the OPERA \cite{OP} and MINOS \cite{MI} collaborations
considered a neutrino beam with the energy peaked at $k=p$, and its
distribution function corresponds to a delta-function 
\begin{equation}
f_k=\delta \left( 1-\frac kp\right)  \label{dis}
\end{equation}
in 1-dimensional momentum space, so that the particle number density (\ref
{n1}) is estimated\textrm{\ }so 
\begin{equation}
n=\frac p\pi  \label{n2}
\end{equation}
and, according to (\ref{e0}) and (\ref{fb}), the energy spectrum is

\begin{equation}
E=\sqrt{p^2+m^2}-\lambda n^b=\sqrt{p^2+m^2}-\frac \lambda {\pi ^b}\,p^b
\label{e0a}
\end{equation}

\section{Vector self-interaction}

Consider more general form of vector self-interaction \cite{DIR2010} 
\begin{equation}
V=\frac \lambda {b+1}\left( \bar \psi \gamma _\nu \psi \bar \psi \gamma ^\nu
\psi \right) ^{b+1}  \label{u1}
\end{equation}
corresponding to the Dirac equation

\begin{equation}
\left( i\gamma ^\mu \partial _\mu -m+F\right) \psi =0  \label{d1}
\end{equation}
where 
\begin{equation}
F=\lambda \left( \bar \psi \gamma _\nu \psi \bar \psi \gamma ^\nu \psi
\right) ^b\gamma _\mu \left( \bar \psi \gamma ^\mu \psi \right)  \label{f1}
\end{equation}
At $b=1$ it is no more than the Heisenberg model of self-interaction \cite
{H57}\textsc{\ } 
\begin{equation}
F=\lambda \gamma _\mu \left( \bar \psi \gamma ^\mu \psi \right)  \label{f111}
\end{equation}

When we consider 1-dimensional neutrino beam, we take into account that
gamma-matrices in (1+1)-dimensional representation \cite{DIR2010} 
\begin{equation}
\gamma ^0=\left( 
\begin{array}{cc}
1 & 0 \\ 
0 & -1
\end{array}
\right) \qquad \gamma ^x=\left( 
\begin{array}{cc}
0 & i \\ 
i & 0
\end{array}
\right)  \label{gam}
\end{equation}
satisfy standard commutation relations 
\begin{equation}
\left\{ \gamma _\mu ,\gamma _\nu \right\} =2\eta _{\mu \nu }\qquad \eta
_{\mu \nu }=\left( 
\begin{array}{cc}
1 & 0 \\ 
0 & -1
\end{array}
\right)  \label{com}
\end{equation}
Substituting stationary plane wave solution\textrm{\ } 
\begin{equation}
\psi =\left( 
\begin{array}{c}
u \\ 
v
\end{array}
\right) \exp \left( ip_xx-iEt\right) \qquad \vec p=\left( p_x,0,0\right)
\label{om1}
\end{equation}
in the Dirac equation (\ref{d2}), we have \cite{DIR2010} \textrm{\ } 
\begin{equation}
\begin{array}{c}
ip_xv=\left( E+\omega -m\right) u \\ 
-ip_xu=\left( E+\omega +m\right) v
\end{array}
\label{co}
\end{equation}
instead of (\ref{dc}), while 
\begin{equation}
\omega =\lambda \left( \left| u\right| ^2+\left| v\right| ^2\right)
^b=\lambda \left( \psi ^{\dagger }\psi \right) ^b  \label{f1b}
\end{equation}
formally coincides with (\ref{f}). The linear system (\ref{co}) has solution
if and only if condition (\ref{e0}) is satisfied. Again, applying the
mean-field approximation 
\begin{equation}
\omega \simeq \omega _{*}=\lambda \left\langle \psi ^{\dagger }\psi
\right\rangle ^b=\lambda n^b  \label{f1c}
\end{equation}
we obtain the same formula for the energy spectrum (\ref{e0a}) of a
1-dimensional neutrino beam.

The group velocity is immediately calculated 
\begin{equation}
v=\frac p{\sqrt{p^2+m^2}}-\frac{\lambda b}{\pi ^b}\,p^{b-1}  \label{v1}
\end{equation}
that tends to 
\begin{equation}
v\simeq 1-\frac{\lambda b}{\pi ^b}p^{b-1}-\frac{m^2}{2p^2}  \label{v1b}
\end{equation}
in the ultra-relativistic limit $E\simeq p\gg m$. The latter term in (\ref
{v1b}) is negligible even at the upper bound of neutrino mass (\ref{mu}),
and superluminal velocity (\ref{del})-(\ref{del2}) can be explained by the
second term if $\lambda b<0$. Indeed, the power energy spectrum (\ref{s}) is
compatible with (\ref{v1b}) if 
\begin{equation}
b=a=0.4\div 1.18  \label{b}
\end{equation}
\begin{equation}
\lambda =-\left( 0.5\div 1.6\right) \times 10^{-4}  \label{la1}
\end{equation}
Particularly, the coupling constant may vary in the range 
\begin{equation}
\lambda =-\left( 0.7\div 0.9\right) \times 10^{-4}  \label{la11}
\end{equation}
when $a=b=1$ that corresponds to the Heisenberg model (\ref{f111}).
Therefore, the origin of superluminal velocity of neutrino \cite{OP,MI} can
be associated with vector self-interaction (\ref{u1}) if the coupling
constants are properly identified (\ref{b})-(\ref{la1}).


\section{Scalar self-interaction}

Consider the Lagrangian (\ref{li}) with scalar self-interaction 
\begin{equation}
V=\frac \lambda {b+1}\left( \bar \psi \psi \right) ^{b+1}  \label{u2}
\end{equation}
The Dirac equation (\ref{d0}) has the form

\begin{equation}
\left( i\gamma ^\mu \partial _\mu -m+W\right) \psi =0  \label{d2}
\end{equation}
where 
\begin{equation}
W=\lambda \left( \bar \psi \psi \right) ^b  \label{f2}
\end{equation}

Substituting stationary wave function $\psi =\varphi \left( \vec r\right)
\exp \left( -iEt\right) $ 
and the effective mass 
\begin{equation}
m_{*}=m-W  \label{m2}
\end{equation}
in (\ref{d2}), we obtain equation 
\begin{equation}
\left( -i\vec \alpha \cdot \nabla +\beta m_{*}\right) \varphi =E\varphi
\label{d2c}
\end{equation}
that has plane-wave solution 
\begin{equation}
\varphi =\left( 
\begin{array}{c}
\phi \\ 
\chi
\end{array}
\right) \exp \left( i\vec p\cdot \vec r\right) \qquad \chi =\frac{\vec
\sigma \cdot \vec p}{E+m_{*}}\phi  \label{om2}
\end{equation}
for a free particle with the energy spectrum 
\begin{equation}
E=\sqrt{\vec p^2+m_{*}^2}  \label{e2}
\end{equation}
and relevant group velocity 
\begin{equation}
v=\frac 1{\sqrt{p^2+m_{*}^2}}\left( p+m_{*}\frac{dm_{*}}{dp}\right)
\label{v2}
\end{equation}
Solution (\ref{om2}) also implies 
\begin{equation}
\bar \psi \psi =\left\| \phi \right\| ^2-\left\| \chi \right\| ^2=\frac{m_{*}%
}E\psi ^{\dagger }\psi  \label{no1}
\end{equation}
where 
\begin{equation}
\psi ^{\dagger }\psi =\left\| \phi \right\| ^2+\left\| \chi \right\| ^2
\label{no2}
\end{equation}

Applying the mean-field approximation, we neglect correlations in (\ref{f2})
and use the effective interaction 
\begin{equation}
W\simeq W_{*}=\lambda n_s^b  \label{f2b}
\end{equation}
where 
\begin{equation}
n_s=\left\langle \bar \psi \psi \right\rangle =\frac{m_{*}}En  \label{ns}
\end{equation}
is the scalar density and $n$ is the particle number density (\ref{n}). For
a many-particle system it is determined by formula 
\begin{equation}
n_s=\frac 2{\left( 2\pi \right) ^q}\int\limits_0^\infty \frac{m_{*}\left(
k\right) }{E\left( k\right) }f_k\,d^qk  \label{ns1}
\end{equation}
The distribution function $f_k$ of a neutrino beam with the energy peaked at 
$k=p$ is presented by a delta-function (\ref{dis}) in 1-dimensional momentum
space, so that the scalar density (\ref{ns}) is estimated 
\begin{equation}
n_s=\frac p\pi \frac{m_{*}\left( p\right) }{E\left( p\right) }  \label{ns2}
\end{equation}
Substituting (\ref{f2b}) and (\ref{ns2}) in (\ref{m2}), we obtain
self-consistent equation for the effective mass 
\begin{equation}
m_{*}\left( p\right) =m-\lambda n_s^b=m-\lambda \left[ \frac p\pi \frac{%
m_{*}\left( p\right) }{\sqrt{\left( p^2+m_{*}^2\right) }}\right] ^b
\label{m2b}
\end{equation}

Consider the ultra--relativistic limit $E\simeq p\gg m_{*}$. The energy (\ref
{e2}) is reduced to 
\begin{equation}
E=p+\frac{m_{*}^2}{2p}=p+\frac{m^2}{2p}-\frac{mW_{*}}p+\frac{W_{*}^2}{2p}
\label{e2b}
\end{equation}
while the effective interaction (\ref{f2b}) is reduced to 
\begin{equation}
W_{*}\simeq \frac \lambda {\pi ^b}m_{*}^b  \label{f2c}
\end{equation}
and velocity (\ref{v2}) is estimated 
\begin{equation}
v\simeq 1-\frac{m_{*}^2}{2p^2}  \label{v2b}
\end{equation}
Although the deviation from the speed of light may be visible at large $%
W_{*} $, the velocity (\ref{v2b}) is subluminal at any choice of constants $%
\lambda $ and $b$. Therefore, the scalar self-interaction cannot be
responsible for the superluminal neutrino velocity.


\section{Scalar-vector self-interaction}

Now consider a mixed variant of scalar-vector self-interaction in the form 
\cite{DIR2007,DIR2008} 
\begin{equation}
V=\lambda \frac{\left( \psi ^{\dagger }\psi \right) ^{b+1}}{\left( \bar \psi
\psi \right) ^b}  \label{u3}
\end{equation}
that corresponds to the Dirac equation

\begin{equation}
\left( i\gamma ^\mu \partial _\mu -m+W+\gamma ^0\omega \right) \psi =0
\label{d3}
\end{equation}
where 
\begin{equation}
\omega =\lambda \left( b+1\right) \frac{\left( \psi ^{\dagger }\psi \right)
^b}{\left( \bar \psi \psi \right) ^b}\qquad W=-\lambda b\frac{\left( \psi
^{\dagger }\psi \right) ^{b+1}}{\left( \bar \psi \psi \right) ^{b+1}}
\label{f3}
\end{equation}

Substituting a stationary plane wave solution 
\begin{equation}
\varphi =\left( 
\begin{array}{c}
\phi \\ 
\chi
\end{array}
\right) \exp \left( i\vec p\cdot \vec r-iEt\right)  \label{om3}
\end{equation}
and the effective mass (\ref{m2}) in (\ref{d3}), we have a linear system of
equations for the spinors 
\begin{equation}
\begin{array}{c}
\vec \sigma \cdot \vec p\chi =\left( E+\omega -m_{*}\right) \phi \\ 
\vec \sigma \cdot \vec p\phi =\left( E+\omega +m_{*}\right) \chi
\end{array}
\label{d1c}
\end{equation}
that has solution 
\begin{equation}
\chi =\frac{\vec \sigma \cdot \vec p}{E+\omega +m_{*}}\phi  \label{om3b}
\end{equation}
if and only if the particles has the energy spectrum 
\begin{equation}
E=\sqrt{m_{*}^2+p^2}-\omega  \label{e3}
\end{equation}
where the effective mass is 
\begin{equation}
m_{*}=m-W  \label{m3}
\end{equation}

According to (\ref{om3}) and (\ref{om3b}) we have 
\begin{equation}
\frac{\psi ^{\dagger }\psi }{\bar \psi \psi }=\frac{\left\| \phi \right\|
^2+\left\| \chi \right\| ^2}{\left\| \phi \right\| ^2-\left\| \chi \right\|
^2}=\frac{\sqrt{m_{*}^2+p^2}}{m_{*}}  \label{no3}
\end{equation}
and the effective mass (\ref{m3}) is determined by self-consistent equation 
\begin{equation}
m_{*}=m-\lambda b\frac{\sqrt{\left( m_{*}^2+p^2\right) ^{b+1}}}{m_{*}^{b+1}}
\label{m3b}
\end{equation}
that is reduced to 
\begin{equation}
m_{*}=m-\lambda b\frac{p^{b+1}}{m_{*}^{b+1}}  \label{m3c}
\end{equation}
in the ultra-relativistic limit $p\gg m_{*}$, so that 
\begin{equation}
\left[ \frac{b+2}{b+1}m_{*}-m\right] m_{*}^{\prime }=\left( m_{*}-m\right) 
\frac{m_{*}}p  \label{m3d}
\end{equation}
The energy spectrum (\ref{e3}) is, then, reduced to 
\begin{equation}
E\simeq p+\frac{m_{*}^2}{2p}-\lambda \left( b+1\right) \frac{p^b}{m_{*}^b}
\label{e3b}
\end{equation}

At large coupling constant $\lambda $, when 
\begin{equation}
m_{*}\gg m  \label{m99a}
\end{equation}
equation (\ref{m3d}) implies 
\begin{equation}
m_{*}^{\prime }=\frac{b+1}{b+2}\frac{m_{*}}p  \label{e99b}
\end{equation}
and, according to equation (\ref{e3b}), the neutrino velocity is subluminal 
\begin{equation}
v\simeq 1-\frac{m_{*}^2}{2p^2}  \label{v99a}
\end{equation}
When 
\begin{equation}
m_{*}\rightarrow m  \label{m66a}
\end{equation}
equation (\ref{m3d}) implies $m_{*}^{\prime }\rightarrow 0$ and, according
to equation (\ref{e3b}), the neutrino velocity is also subluminal 
\begin{equation}
v\simeq 1-\frac{m^2}{2p^2}  \label{v66a}
\end{equation}
When 
\begin{equation}
m_{*}\ll m  \label{m33a}
\end{equation}
equation (\ref{m3d}) implies 
\begin{equation}
m_{*}^{\prime }\simeq \frac{m_{*}}p  \label{e33b}
\end{equation}
and, according to equation (\ref{e3b}), the neutrino velocity is
superluminal 
\begin{equation}
v-1\simeq \frac{m_{*}^2}{2p^2}\ll \frac{m^2}{2p^2}  \label{v33a}
\end{equation}
but it negligible with respect to estimation (\ref{vv}) for a free massive
neutrino and, hence, cannot achieve velocity (\ref{del})-(\ref{del2}) that
was observed in experiments \cite{OP,MI}.

\section{Conclusion}

The model of neutrino with self-interaction can explain superluminal
velocity (\ref{del})-(\ref{del2}) recorded by the OPERA and MINOS\
collaborations \cite{OP,MI}. When the Lagrangian includes vector
self-interaction in the form of (\ref{u}) or (\ref{u1}), the energy spectrum
of a 1-dimensional neutrino beam (\ref{e0a}) includes additional term (\ref
{f1b}), and the neutrino can move faster than light. Its energy spectrum is
compatible with the power energy spectrum (\ref{s}), and the neutrino
velocity is associated with the observed data when the coupling constants
satisfy constraints (\ref{b}) and (\ref{la1}). The Heisenberg model of
self-interacting spin-1/2 particle (\ref{f111}) can also give satisfactory
result under appropriate choice of the coupling constant (\ref{la1})

\textrm{\ }Neutrino with scalar self-interaction (\ref{u2}) is always
subluminal (\ref{v2b}). Scalar-vector self-interaction (\ref{u3}) can result
in the group velocity (\ref{v33a}) above the speed of light but it is much
smaller than the observed values (\ref{del})-(\ref{del2}).

Although the only model of vector self-interaction (\ref{u1}) is suitable
for explanation of superluminal neutrino, it looks very promising because
this effect is universal and does not depend on the medium or external
fields. It should be emphasized that the neutrino mass is not important
here, and the observed superluminal velocity can be produced by a massless
particle. We have applied the mean-field approximation (\ref{f1c}) for
obtaining the effective energy spectrum (\ref{e0}) of a plane-wave solution (%
\ref{om1}). It is enough for estimation of the neutrino velocity. Exact
analysis of equation (\ref{co}) may concern the problem of neutrino flavor
oscillations that is the subject for further research.

The author is grateful to Erwin Schmidt for discussions.

\end{document}